\documentclass[pra,reprint,superscriptaddress,amsmath,amssymb,aps,showpacs, floatfix]{revtex4-1}

\usepackage{amsmath}
\usepackage{amssymb}
\usepackage{graphicx}
\usepackage[utf8]{inputenc}
\usepackage{epstopdf}
\usepackage{booktabs}
\usepackage{color}
\usepackage{xcolor}
\usepackage[percent]{overpic}
\usepackage{verbatim}
\newcommand{\ba}{\begin{eqnarray}}
\newcommand{\ea}{\end{eqnarray}}
\newcommand{\be}{\begin{equation}}
\newcommand{\ee}{\end{equation}}
\newcommand{\bea}{\begin{eqnarray}}
\newcommand{\eea}{\end{eqnarray}}

\newcommand{\bra}[1]{\langle#1\rvert}
\newcommand{\ket}[1]{\lvert#1\rangle}

\usepackage{theorem}

\theorembodyfont{\rmfamily}

\def\openone{\leavevmode\hbox{\small1\kern-3.3pt\normalsize1}}

\begin{document}
\title{Optimal Phonon-to-Spin Mapping in a system of a trapped ion}
\author{Matthias M. Müller}
\affiliation{Center for Integrated Quantum Science and Technology, Institute for Complex Quantum Systems, University of Ulm, Albert-Einstein-Allee 11, D-89069 Ulm, Germany}
\author{Ulrich G. Poschinger}
\affiliation{QUANTUM, Institut für Physik, Universität Mainz, D-55128 Mainz, Germany}
\author{Tommaso Calarco}
\author{Simone Montangero}
\affiliation{Center for Integrated Quantum Science and Technology, Institute for Complex Quantum Systems, University of Ulm, Albert-Einstein-Allee 11, D-89069 Ulm, Germany}
\author{Ferdinand Schmidt-Kaler}
\affiliation{QUANTUM, Institut für Physik, Universität Mainz, D-55128 Mainz, Germany}

\date{\today}
\pacs{02.60.Pn,03.65.Aa,03.65.Wj,32.80.Qk}
%02.60.Pn Numerical optimization
%03.65.Aa Quantum systems with finite Hilbert space
%03.65.Wj State reconstruction, quantum tomography
%32.80.Qk Coherent control of atomic interactions with photons

\begin{abstract}
We propose a protocol for measurement of the phonon number distribution of a harmonic oscillator based on selective mapping to a discrete spin-1/2 degree of freedom. We consider a system of a harmonically trapped ion, where a transition between two long lived states can be driven with resolved motional sidebands.
The required unitary transforms are generated by amplitude-modulated polychromatic radiation
fields, where the time-domain ramps
are obtained from numerical optimization by application of the Chopped RAndom Basis (CRAB) algorithm. We provide a detailed analysis of the scaling behavior of the attainable fidelities and required times for the mapping transform with respect to the size of the Hilbert space. %, and compare the control scheme to a na{\"i}ve approach with a constant monochromatic drive field. The latter leads to similar scaling as for the Poincaré recurrence problem.
As one application we show how the mapping can be employed as a building block for experiments which require measurement of the work distribution of a quantum process.
\end{abstract}

\maketitle
\section{Introduction}

Trapped ions represent a system where continuous and discrete degrees of freedom can be jointly manipulated and measured in the quantum regime \cite{Gardiner,Kneer,Leibfried2003}. Continuous degrees of freedom are given by oscillatory motion in the trap potential, while the discrete ones are given by internal (spin) states. For coherent spin manipulations, optical or microwave radiation can be employed. This enables a wide variety of applications in the fields of quantum computing \cite{CiracZollerGate,Blatt2008}, quantum simulation \cite{Gerritsma,KimMonroe} and quantum metrology \cite{Rosenband}. 
In this work we propose a scheme that enables single-shot interrogation of the harmonic oscillator degree of freedom and thus paves the way for novel applications of trapped ions.
This is motivated by the persisting difficulty to implement single-shot and/or quantum-non-demolition (QND) measurements of the phonon number for trapped ions, which is mainly due to the predominant harmonicity of the trap potential. Ref. \cite{Huber} proposes a QND filtering scheme, while Ref. \cite{An} demonstrates a single-shot readout. Both schemes require multiple iterations of unitary manipulation and spin readout, which is time consuming and imposes significant experimental complexity.
While we consider  the paradigmatic system of trapped laser-cooled ions, the scheme is applicable 
to other systems such as atoms in cavities \cite{LawEberly, Rojan} or superconducting qubits \cite{Merkel}.
It is therefore of general interest to study the quantum \textit{controllability} of this class of systems in detail and under the assumption of realistic parameters.

In this work, we exploit the controllability of the system to construct a selective, unitary \textit{phonon-to-spin} mapping scheme, where the spin degree of freedom undergoes a flip operation conditioned on the phonon number of the motional mode. Our scheme is fully unitary, however it does not directly serve as a single-shot phonon number measurement, it rather yields a dichotomic result - namely whether the system is found in a specified number state or not. This is a consequence of the fact that the spin-1/2 degree-of-freedom is probed.
We show how this can be enhanced to yield a QND measurement scheme.

The controllability of the problem has been analyzed by \cite{Rangan}, and the control problem has been tackled for state-to-state transfer \cite{LawEberly} and gate optimization \cite{Yuan, Mischuck}.
\begin{figure}[htp!]
\begin{overpic}[width=0.45\textwidth]{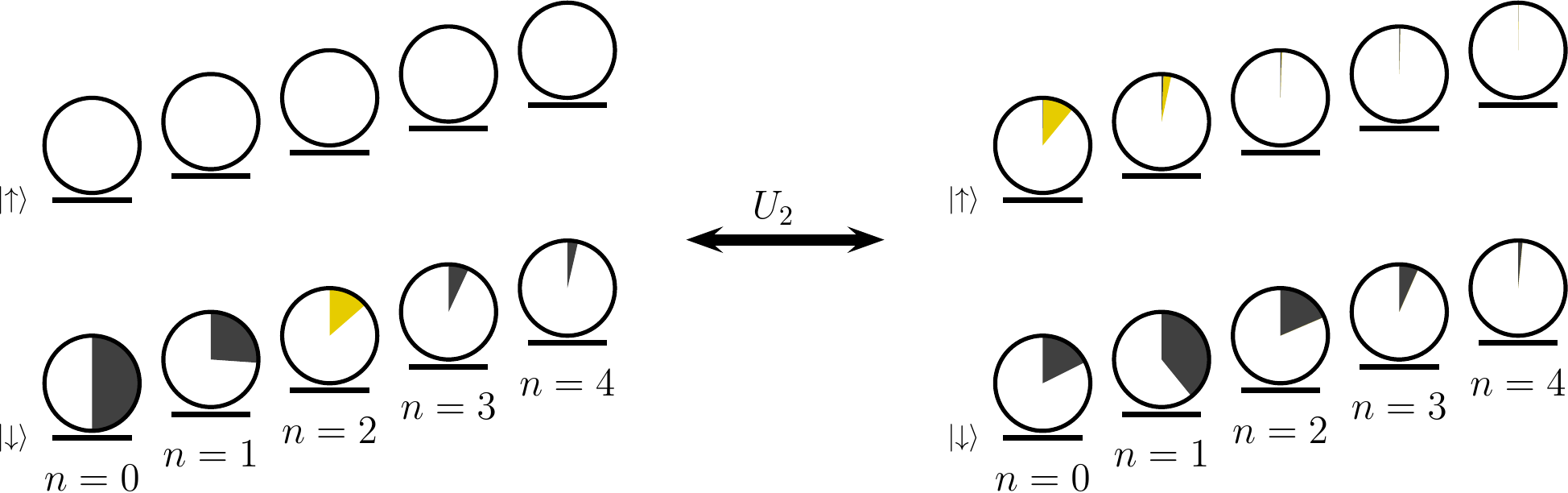}
 \put(-1,17){\color{white}\rule{8pt}{10pt}}
\put(-4,18){\color{black}$|\uparrow\rangle$}
 \put(59,17){\color{white}\rule{8pt}{10pt}}
\put(57,18){\color{black}$|\uparrow\rangle$}
 \put(-1,2){\color{white}\rule{8pt}{10pt}}
\put(-4,2){\color{black}$|\downarrow\rangle$}
 \put(59,2){\color{white}\rule{8pt}{10pt}}
\put(57,2){\color{black}$|\downarrow\rangle$}
 \put(48,17){\color{white}\rule{8pt}{10pt}}
\put(48,18){\color{black}$U_m$}
\put(45,11){\color{black}$m=2$}
\end{overpic}
 \caption{Schematic depiction of the spin mapping $U_m$ for the case of \mbox{$m=2$} (initial state left, final state right): Starting from a product state $\sum_n c_n |\downarrow,n\rangle$ with spin down and occupation in different number states of the harmonic oscillator (the circle diagrams show $|c_n|^2$), only the occupation in the number state $|n=m=2\rangle$ ($|c_2|^2$, yellow) shall be excited to $\ket{\uparrow}$ while the occupation initially in number states $\ket{n\neq 2}$ ($\sum_{n\neq 2} |c_n|^2$, gray) shall end up in $\ket{\downarrow}$. As shown on the right after the mapping $|c_2|^2$ (yellow) is distributed over the spin up states while $\sum_{n\neq 2} |c_n|^2$ (gray) is distributed over the spin down states. Note that only the spin population of the final state matters while the state of the motional mode is not specified.}\label{fig:phononfilter}
\end{figure}
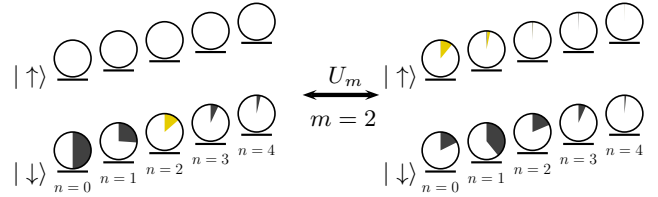
This work is an example for \textit{incomplete control}: We seek to map a specific vibrational level $m\,\in\, 0\dots N-1$ on a spin state, see Fig.~\ref{fig:phononfilter}. In other words, we construct a unitary transformation which flips the spin only for a predefined vibrational number state $m$ and leaves the spin unaffected for the remaining levels $n\neq m$. However, the state of the motional degree of freedom for the final state, is not specified.
This allows us to fix only $N$ parameters (specifying the final time spins conditional the initial harmonic oscillator state) of the full $4N^2-1$ dimensional unitary, thus reducing the complexity of the control problem and allowing for higher fidelities in a shorter time. In this sense the control is incomplete and we can treat systems with larger $N$.

Already for small sizes of the respective Hilbert space no intuitive solution for this control problem is available. We therefore construct the required control fields numerically, based on the Chopped Random Basis (CRAB) \cite{Doria,Caneva} optimization algorithm.

In this work, we first outline the mapping scheme, and integrate this mapping into a quantum non demolition filter in section \ref{sec:Application}. In section \ref{sec:system} we will describe in detail the system of the trapped ion that we want to use to realize the filter, and define the fidelity measure of the mapping. Section \ref{sec:simulation} will introduce the numerical method and parameters for simulation of the system and section \ref{sec:optimization} will specify the optimization method and present results from optimization.

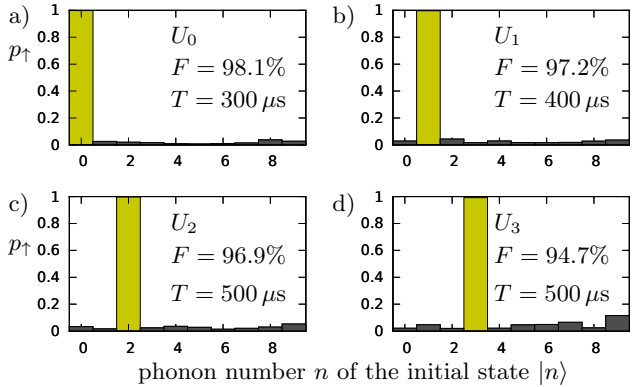
\begin{figure}
\begin{overpic}[width=0.48\textwidth]{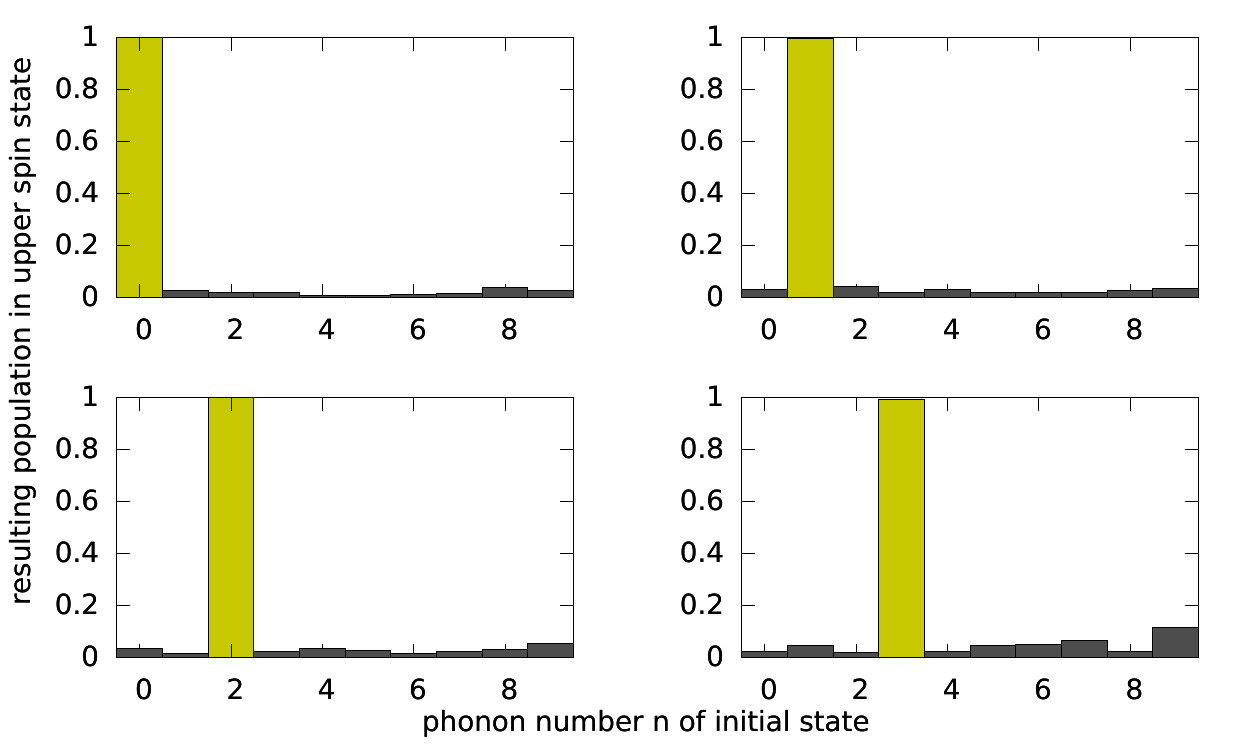}
\put(3,0){\color{white}\rule{200pt}{9pt}}
\put(20,0){\color{black}phonon number $n$ of the initial state $|n\rangle$}
\put(0,10){\color{white}\rule{8pt}{120pt}}
\put(0,50){\color{black}$p_{\uparrow}$}
\put(0,20){\color{black}$p_{\uparrow}$}
\put(25,52){\color{black}$U_0$}
\put(25,47){\color{black}$F=98.1\%$}
\put(25,42){\color{black}$T=300\,\mathrm{\mu s}$}
%\put(5,25){\color{white}\rule{15pt}{95pt}}
\put(25,23){\color{black}$U_2$}
\put(25,18){\color{black}$F=96.9\%$}
\put(25,12){\color{black}$T=500\,\mathrm{\mu s}$}
\put(75,52){\color{black}$U_1$}
\put(75,47){\color{black}$F=97.2\%$}
\put(75,42){\color{black}$T=400\,\mathrm{\mu s}$}
\put(75,23){\color{black}$U_3$}
\put(75,18){\color{black}$F=94.7\%$}
\put(75,12){\color{black}$T=500\,\mathrm{\mu s}$}
\put(0,55){a)}
\put(50,55){b)}
\put(0,26){c)}
\put(50,26){d)}
\end{overpic}
\caption{Performance of $U_m(T)$ for different $m$ as it acts on the states $|\downarrow,n\rangle$. a) $m=0$, b) $m=1$, c) $m=2$,  d) $m=3$. The columns show the resulting population in the upper spin state $p_{\uparrow}=\sum_k |\langle\uparrow,k |U_m(T)|n,\downarrow\rangle|^2$ depending on the initial phonon number state $|n\rangle$: the yellow colums show the population for $m$, where the spin should be flipped, the gray columns show the spin up population for phonon numbers $n$, where we want to keep the state in spin down.}\label{fig:filtersketch}
\end{figure}

\subsection{The Mapping Scheme}
The purpose is to map a specific vibrational level $\ket{m}$ on a spin degree of freedom, see Fig.~\ref{fig:phononfilter}. We start with a product state $\rho(0)=\ket{\downarrow}\otimes \rho^{\mathrm{HO}}(0)$, where the spin is initialized to $\ket{\downarrow}$ and the harmonic oscillator mode is in an arbitrary pure or mixed state characterized by $\rho^{\mathrm{HO}}(0)$. By applying a pulse sequence of up to three radiation fields we create a unitary transformation $U_m$, that flips the spin for initial occupation in the vibrational level $\ket{m}$ and keeps spin down for $\ket{n}\neq \ket{m}$. That is, $U_m$ maps 

\begin{eqnarray}
\ket{\downarrow}\otimes|n\rangle
\stackrel{U_m}{\longleftrightarrow}
\begin{cases}
\ket{\uparrow}\otimes\sum_{n'} c^{(n,m)}_{n'}|n'\rangle\quad n=m\\
%%%
\ket{\downarrow}\otimes\sum_{n'} c^{(n,m)}_{n'}|n'\rangle\quad n\neq m
\end{cases}\,.
\end{eqnarray}

We will distinguish the ideal map $U_m$ that we seek to implement from the actual finite-time evolution $U_m(T)$.
Fig.~\ref{fig:filtersketch} shows the sucess of such maps for $m=0,1,2,3,4$ as obtained by CRAB. The details of the results will be explained in section \ref{sec:optimization}.

Note that for each initial state $\ket{n}$, the vibrational mode is supposed to be in an unspecified superposition state after the filter unitary transform. Thus, only direct readout or coherent spin manipulations are to be applied after the filter operation.
Using trapped cold ions, the spin can be read out with fidelities of 99.99\,\% \cite{Myerson}, and preparation of identical input states is current state of the art. We can repeat a sequence of preparation, mapping transform $U_m(T)$ and spin readout to obtain a statistical estimate for $p_m=\rho_{m,m}^{\mathrm{HO}}$, the initial occupation in $\ket{m}$. Performing this for all number states in a truncated subspace yields the phonon distribution. With additional displacement  operations after preparation, this can be extented to a full tomography scheme \cite{Leibfried1996}.

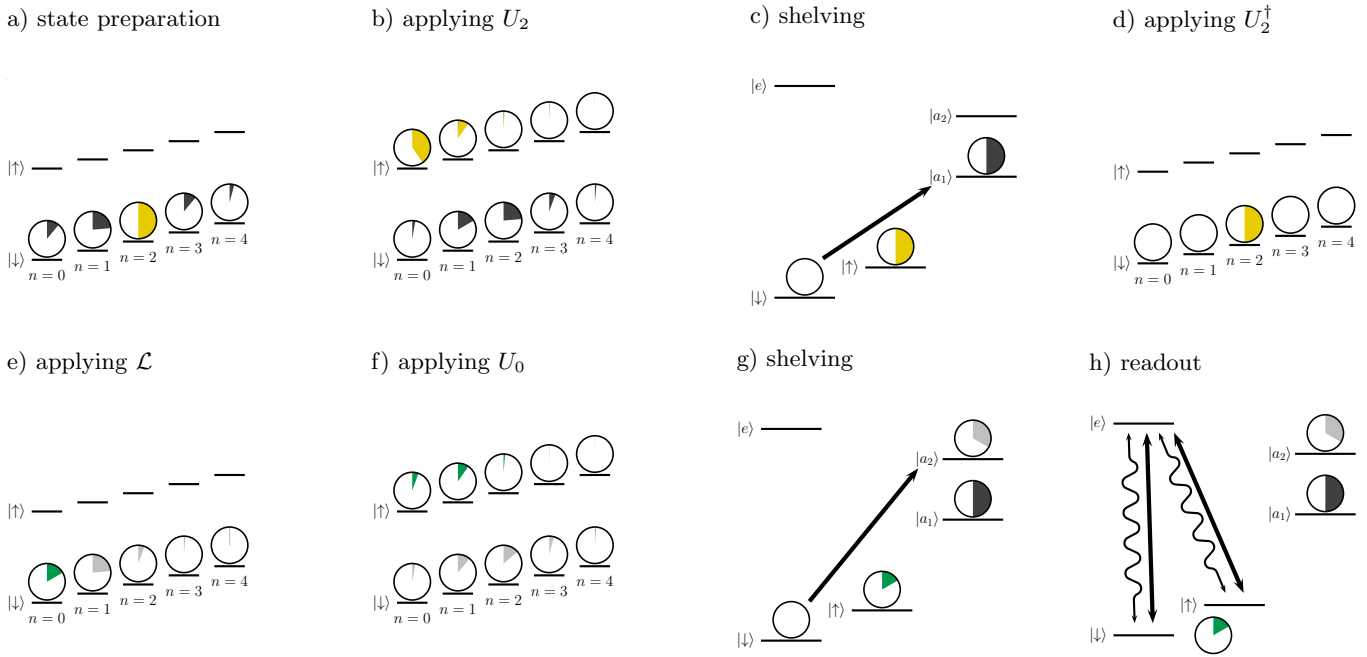
\begin{figure*}
\vspace*{0.5cm}
 \begin{minipage}{0.5\textwidth}\flushleft
 \begin{overpic}[width=0.9\textwidth]{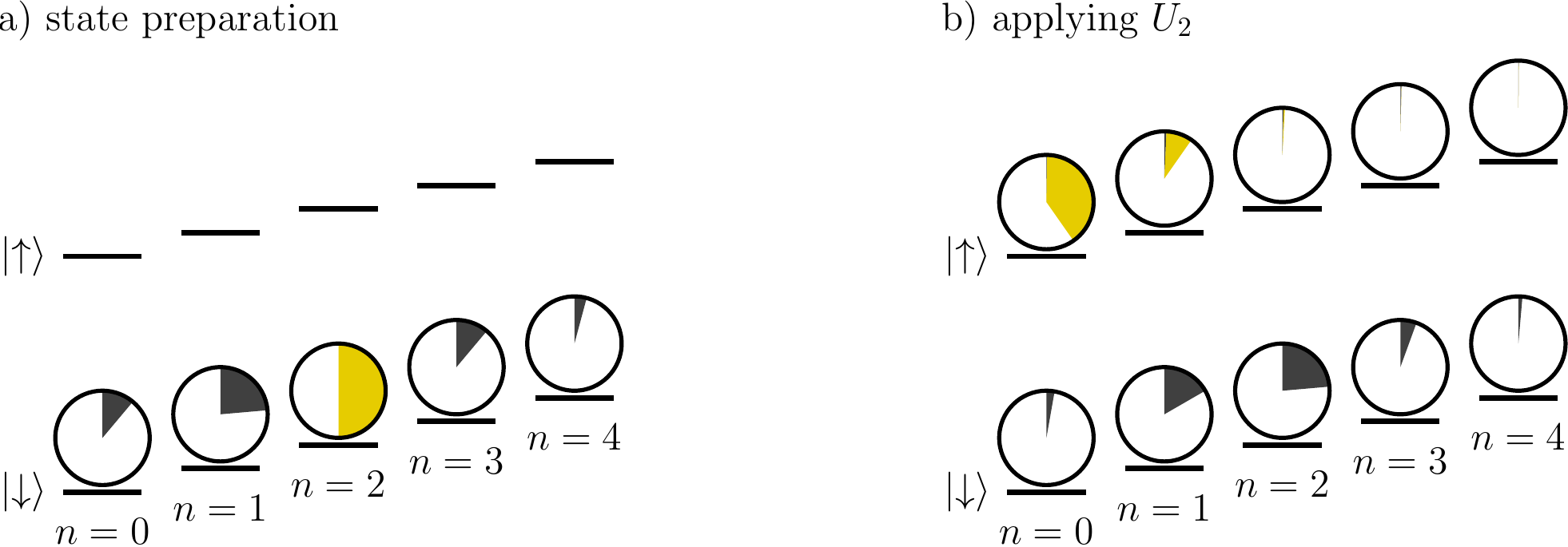}
\put(0,32){\color{white}\rule{80pt}{10pt}}
\put(0,42){\color{black}a) state preparation}
\put(50,32){\color{white}\rule{80pt}{10pt}}
\put(60,42){\color{black}b) applying $U_2$}
\end{overpic}
\end{minipage}%
 \begin{minipage}{0.04\textwidth}
  \qquad
 \end{minipage}%
 \begin{minipage}{0.22\textwidth}\centering
 \begin{overpic}[width=0.9\textwidth]{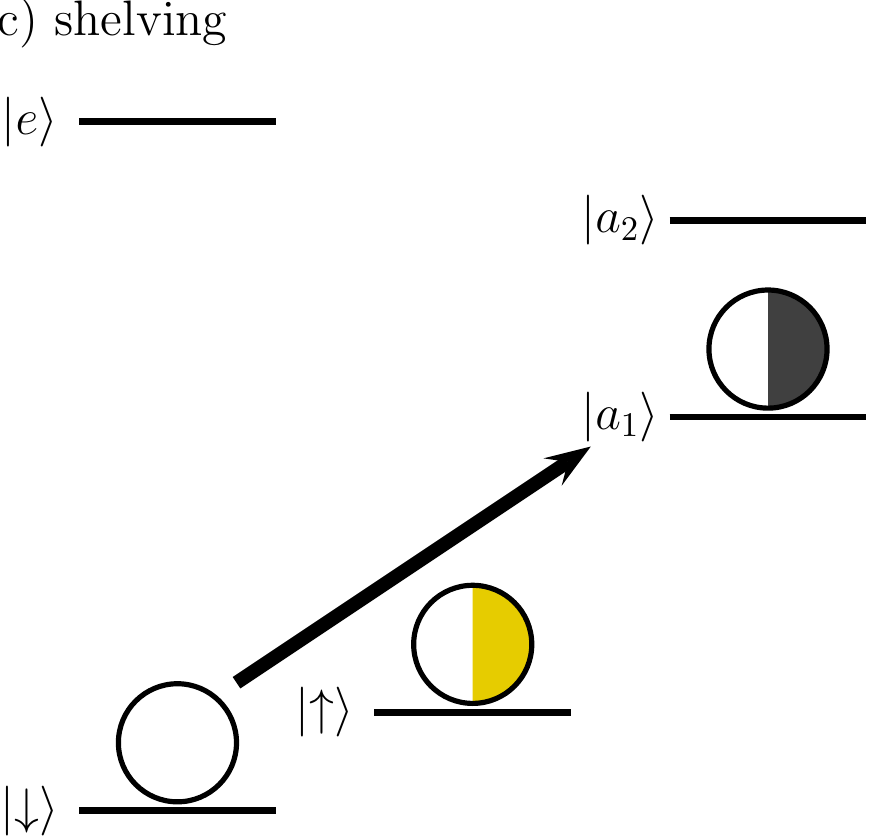}
  \put(-1,89){\color{white}\rule{80pt}{10pt}}
\put(0,106){\color{black}c) shelving}    
 \end{overpic}
 \end{minipage}%
 \begin{minipage}{0.04\textwidth}
  \qquad
 \end{minipage}%
 \begin{minipage}{0.2\textwidth}\flushright
 \begin{overpic}[width=0.9\textwidth]{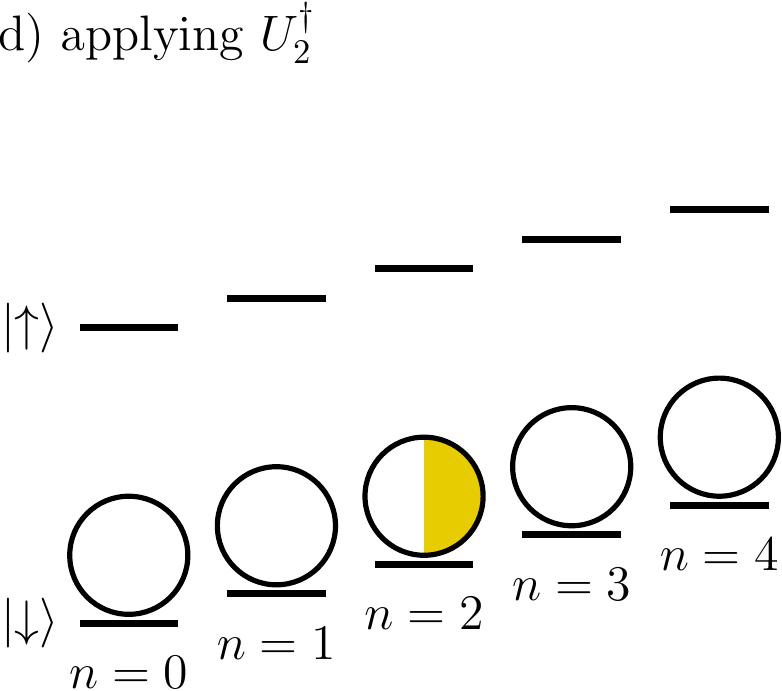}
  \put(-1,69){\color{white}\rule{80pt}{110pt}}
\put(0,106){\color{black}d) applying $U_2^\dagger$}
 \end{overpic}
 \end{minipage}
\vspace*{1cm}\\
%%%%%%%%%%%%%%%%%%%%%%%%%%%%%%%%%%%%%%% Zweite Reihe
 \begin{minipage}{0.5\textwidth}\flushleft
 \begin{overpic}[width=0.9\textwidth]{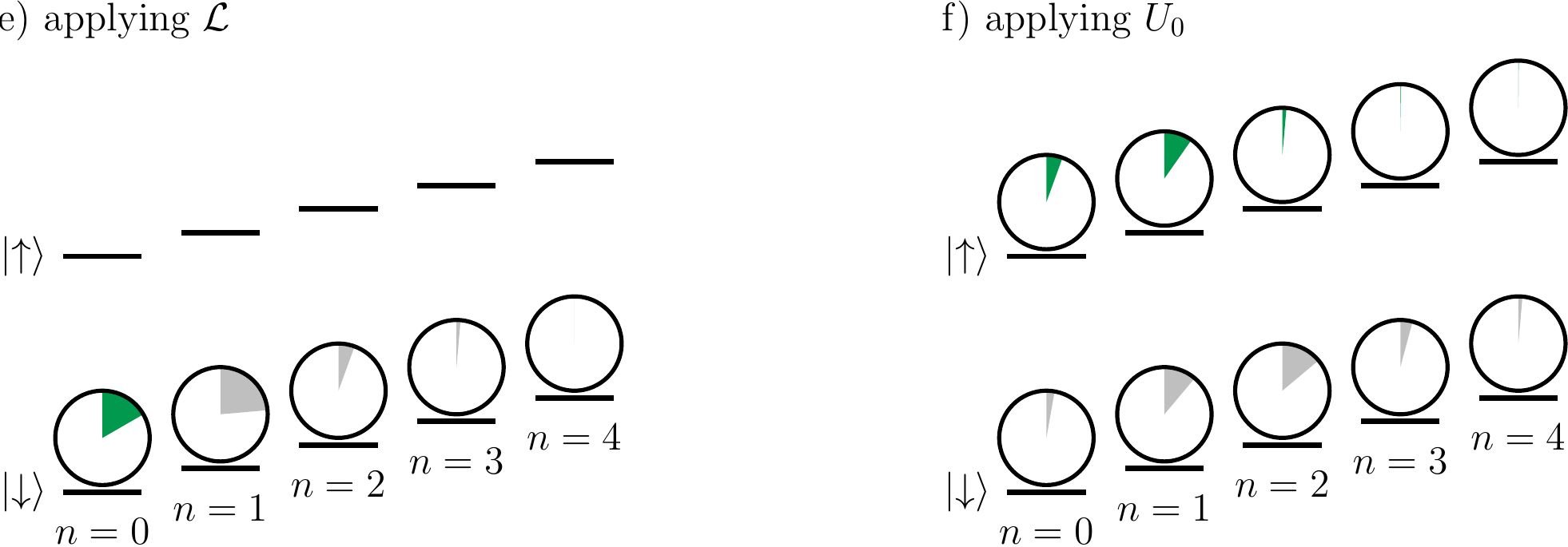}
  \put(-1,32){\color{white}\rule{80pt}{10pt}}
  \put(49,32){\color{white}\rule{80pt}{10pt}}
 \put(0,42){\color{black}e) applying $\mathcal{L}$}
\put(60,42){\color{black}f) applying $U_0$}
 \end{overpic}
 \end{minipage}%
 \begin{minipage}{0.03\textwidth}
  \qquad
 \end{minipage}%
 \begin{minipage}{0.22\textwidth}\centering
 \begin{overpic}[width=0.9\textwidth]{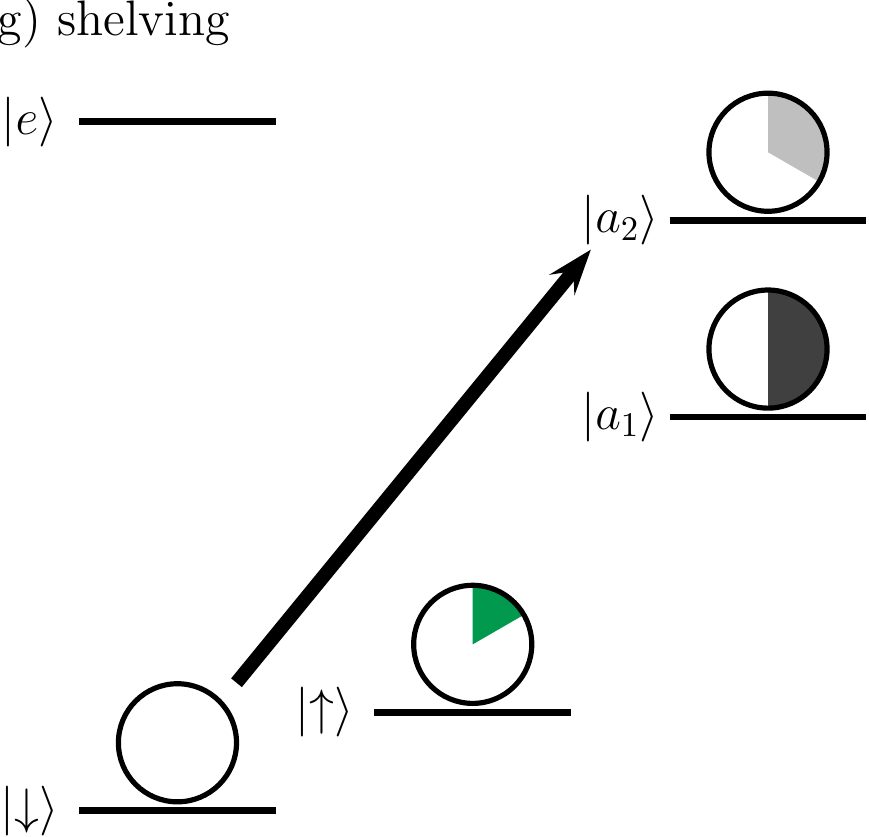}
  \put(-1,90){\color{white}\rule{80pt}{10pt}}
\put(0,106){\color{black}g) shelving}
 \end{overpic}
 \end{minipage}%
 \begin{minipage}{0.03\textwidth}
  \qquad
 \end{minipage}%
 \begin{minipage}{0.22\textwidth}\flushright
 \begin{overpic}[width=0.9\textwidth]{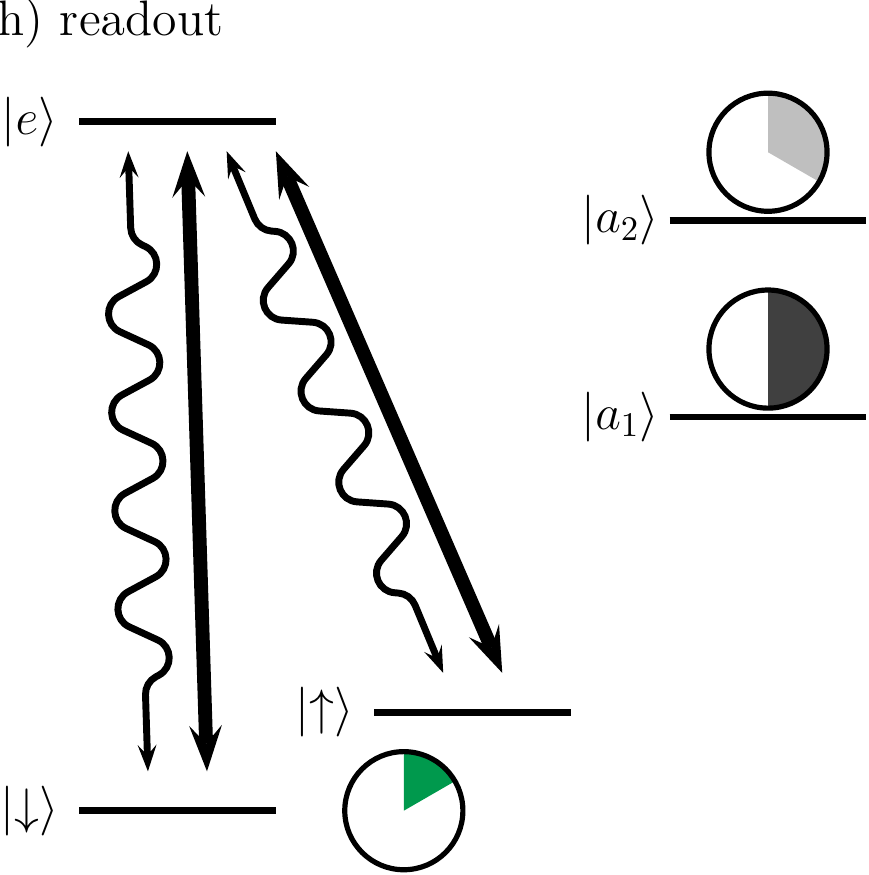}
  \put(-1,90){\color{white}\rule{80pt}{10pt}}
\put(0,106){\color{black}h) readout}
 \end{overpic}
 \end{minipage}
\caption{Employing unitary phonon to spin mapping for a QND filter: The panels show a sequence of mapping and spectroscopic decoupling operations which enables measurement of the work distribution of a quantum process. \textbf{a)} shows a state prepared in $\ket{\downarrow}$ with an arbitrary state of the vibrational mode. \textbf{b)} shows how population from $\ket{m=2}$ is transferred to $\ket{\uparrow}$ by means of $U_2$. The vibrational populations are reshuffled in an unspecified way. In \textbf{c)}, population from $\ket{\downarrow}$ is shelved to a metastable auxiliary state $\ket{a_1}$ for spectroscopic decoupling. \textbf{d)} shows how the population which was not shelved is transferred to $\ket{\downarrow,2}$ by applying the time reversed mapping $U_2^{\dagger}$. In \textbf{e)}, a general quantum process $\mathcal{L}$ (unitary and/or dissipative) takes place which performs work on or transfers heat to the vibrational mode. In \textbf{f)}, the population of $\ket{m'=0}$ is mapped to $\ket{\uparrow}$ 
by means of $U_0$, and the remaining population in $\ket{\downarrow}$ is shelved to another metastable state $\ket{a_2}$ in \textbf{g)}. Finally, $\textbf{h)}$ shows how readout takes place by cycling to population of the ground state through an excited state $\ket{e}$ and detecting the resonance fluorescence.}
\label{fig:filterscheme}
\end{figure*}

\section{Application for filtering and the measurement of work distributions}\label{sec:Application}
In this section, we outline how the unitary mapping operation can be used in conjunction with spectroscopic decoupling \cite{Schindler} to realize a filtering protocol similar to the QND filter proposed in \cite{Davidovich,Huber}. In that scheme, pulses of defined area on a motional sideband realize a low-precision mapping operation. After one mapping operation, the spin is projectively measured by fluorescence detection. Absence of fluorescence indicates that the system is likely to be in $\ket{m}$, and the discrimination efficiency increases upon iterative repetition. This corresponds to preparation in level $\ket{m}$ by QND measurement. For a quantum process acting on the vibrational mode, this scheme allows for measurement of the work distribution by QND preparation of level
$\ket{m}$, the subsequent quantum process $\mathcal{L}$ and a second filtering operation for level $\ket{m'}$. This yields the probability that the quantum process transfers level $\ket{m}$ to $\ket{m'}$.

While this opens up the prospect of studying the non-equilibrium thermodynamics of a well-controlled quantum system, the approach is difficult to implement experimentally. The main reason lies in the fact that a Fock state needs to be preserved over a comparatively long time span, where the repeated unitary driving and dissipative measurement operations take place.
The filtering scheme presented in Ref. \cite{An} works by iterative adiabatic phonon removal, where the event that the system has initially been in Fock state $\ket{n}$ is heralded by bright detection after $n-1$ previous cycles of phonon removal and dark detection. The scheme however requires the deterministic preparation of state $\ket{n}$ after the final bright event, e.g. by convential ladder climbing schemes \cite{Meekhof, Ziesel}.
By contrast, our scheme consists of a fully unitary mapping operation, which can be enhanced to a measurement protocol for work distributions by population transfer from a specific spin state to additional meta-stable levels.

The scheme is explained in detail by Fig. \ref{fig:filterscheme}. 
We start out with an initial state $\ket{\downarrow}\otimes \rho^{\mathrm{HO}}(0)$. We might be interested in the population of some vibrational level $\ket{m}$ (panel a), $m=2$). We apply the mapping $U_m$ to transfer this population to $\ket{\uparrow}$ (b), and the population remaining in $\ket{\downarrow}$, i.e. for all $\ket{n\neq m}$, is transferred to the metastable state $\ket{a_1}$ (c). We then apply the time reversed filter $U_m^{\dagger}$ and obtain the state $\ket{\downarrow,m}$, ignoring the shelved population (d). We then carry out the generic quantum process $\mathcal{L}$ which is to be analyzed, which acts only on the motional degree of freedom (e). The redistribution of population among the Fock states is then analyzed by applying the mapping operation $U_{m'}$ (f) and shelving the population left over in $\ket{\downarrow}$ to a second metastable state $\ket{a_2}$ (g). We can now measure the groundstate population by detection of state-dependent resonance fluorescence by driving a cycling transition to a short-lived excited state $\ket{e}$ (h). For a quantum process 
$\rho'=\mathcal{L}(\rho)$, the measurement sequence yields fluorescence with probability 
\be
P_{f}= \langle m|\rho^{\mathrm{HO}}(0)|m\rangle \left\langle m'|\mathcal{L}(\ket{m}\bra{m})|m'\right\rangle.
\ee
The initial populations can be measured separately by simply performing the fluorescence detection after the first shelving operation. Thus, the measurement scheme yields the transfer matrix of the quantum process $\mathcal{L}$ and therefore its work distribution.\\
The scheme can be implemented with all ion species commonly used in experiments. For $^{40}$Ca$^+$ or  $^{88}$Sr$^+$, the states $\ket{\uparrow},\ket{\downarrow}$ can be identified with the Zeeman sublevels of the $S_{1/2}$ ground state, the auxiliary states $\ket{a_1},\ket{a_2}$ with different sublevels of the $D_{5/2}$ state, which has a lifetime of about $1\,$ms, while the excited state would be the $P_{1/2}$ state.

\section{The System}\label{sec:system}
\begin{figure}
 \begin{minipage}{0.25\textwidth}
 \begin{overpic}[width=0.95\textwidth]{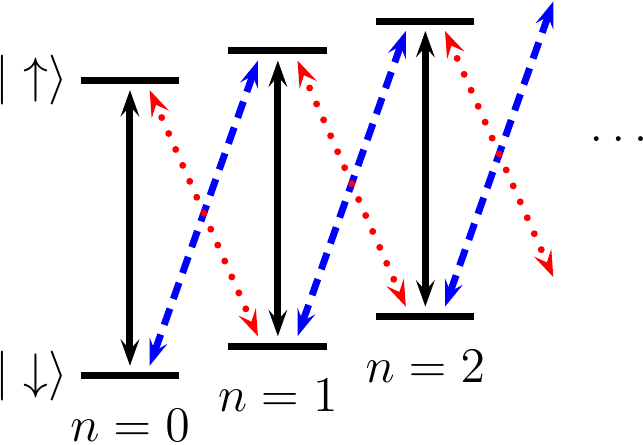}
  \put(-2,71){a)}
 \end{overpic}
 \end{minipage}%
 \begin{minipage}{0.01\textwidth} \qquad
 \end{minipage}%
 \begin{minipage}{0.23\textwidth}\hspace*{0.055\textwidth}
\begin{overpic}[width=1\textwidth]{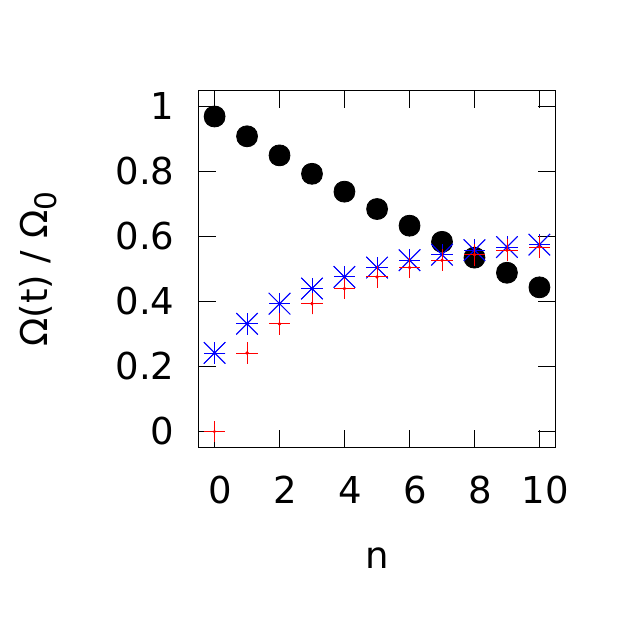}
\put(3,3){\color{white}\rule{100pt}{12pt}}
\put(50,6){\color{black}$n$}
\put(3,10){\color{white}\rule{12pt}{120pt}}
\put(-9,70){\color{black}$M_{n,n}$}
\put(-9,55){\color{blue}$M_{n,n+1}$}
\put(-9,40){\color{red}$M_{n,n-1}$}
\put(-2,87){b)}
\end{overpic}
 \end{minipage}
\caption{a) Energy level scheme of a single trapped ion. The relevant levels are product states of the spin state and the harmonic oscillator modes given by the phonon number states $\ket{n}$. Lasers drive transitions on the carrier as well as on the blue and red sidebands. b) Relative Rabi frequencies (given by the matrix elements $M_{n,n+\delta n}=\Omega^{(\alpha)}_{n,n+\delta n}(t)/\Omega^{(\alpha)}(t)$) for $\eta=0.25$. On resonance this is the relative coupling strength of the carrier transition (black circles, $\delta n=0$), the blue (stars) one for the blue sideband ($\delta n=1$), and the red (crosses) one for the red sideband ($\delta n=-1$).}\label{fig:overview-sketch}
\end{figure}
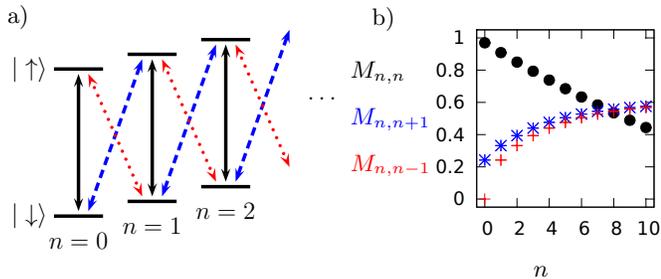

We consider a trapped ion of mass $M$ and limit ourselves to a 1D harmonic oscillator of frequency $\omega_z$. The internal states are coupled by radiation, which is characterized by the Rabi frequency $\Omega$. A sufficiently short wavelength $\lambda$ serves to couple internal and vibrational levels. This condition is described by the Lamb-Dicke parameter $\eta=\frac{1}{\lambda}\sqrt{\frac{\hbar}{2 M\omega_z}}$, specifying the ratio of the ground state wavepacket size to the wavelength of the driving radiation \cite{Leibfried}.

Any ion species which features a closed transition for laser cooling and a qubit transition is suitable. Thus, optical \cite{Schindler}, stimulated Raman \cite{Poschinger} and microwave qubits \cite{Ospelkaus} can be employed.  
The Hilbert space of the system is a product space of the two relevant spin and the vibrational (harmonic oscillator) degrees of freedom. Up to three radiation fields are applied to drive transitions between these levels: One resonant with the carrier transition which connects the two spin states for equal phonon number, and fields driving the blue (red) sidebands. The combined phonon and spin level scheme as well as the laser driven transitions and the dependence of the relative transition strength on the vibrational excitation $n$ are depicted in Fig.~\ref{fig:overview-sketch} for the parameters given in the following section.

\subsection{Hamiltonian}
Three radiation fields contribute to the Hamiltonian. We label the fields by $\alpha=c,b,r$ and characterize them by the bare Rabi frequency $\Omega_0^{(\alpha)}$ and the detuning from the carrier transition $\delta^{(\alpha)}_c=\omega_\alpha-\omega_c$ ($\omega_c$ the carrier transition and $\omega_\alpha$ the frequency of the field). To have the fields on resonance with the carrier ($\Omega_0^{(c)}$) and the blue and red sidebands ($\Omega_0^{(b)}$ and $\Omega_0^{(r)}$, respectively) we have to choose $\delta^{(c)}_c=0$, $\delta^{(b)}_c=\omega_z$, and 
$\delta^{(r)}_c=-\omega_z$, where the trap frequency $\omega_z$ is the energy of one phononic excitation.

Each radiation field does not only resonantly drive transitions, it also causes energy level shifts via off-resonant driving. We neglect terms rotating at $2\omega_z$ or faster, and terms which scale with a $|\delta n|> 1$ matrix element. The Hamiltonian, including resonant and off-resonant effects,  then reads:
\begin{widetext}
\begin{eqnarray}\label{eq:Ham-allfields}
H(t)=\vert \uparrow \rangle\langle \downarrow \vert\otimes\sum_{n=0}\Bigg[
&\sum_{\delta n=-1,0,1}&\frac{\Omega^{(c)}_{n,n+\delta n}(t)}{2}\mathrm{e}^{-\delta n i\omega_z t}\vert n+\delta n \rangle\langle n\vert\nonumber\\
&\sum_{\delta n=0,1\;\;\;\;}&\frac{\Omega^{(b)}_{n,n+\delta n}(t)}{2}\mathrm{e}^{(1-\delta n) i\omega_z t}\vert n+\delta n \rangle\langle n\vert \nonumber\\
&\sum_{\delta n=-1,0\;}&\frac{\Omega^{(r)}_{n,n+\delta n}(t)}{2}\mathrm{e}^{-(1+\delta n) i\omega_z t}\vert n+\delta n \rangle\langle n\vert\Bigg] + h.c.
\end{eqnarray}
\end{widetext}

 The Rabi frequencies depend on the phonon number $n$ and the Lamb-Dicke factor $\eta$ as 
\begin{eqnarray}
 \Omega^{(\alpha)}_{n,n+\delta n}(t)=\Omega^{(\alpha)}(t) M_{n,n+\delta n}(\eta)
\end{eqnarray}
where the matrix elements are modified Laguerre polynomials and are given in the appendix. This dependence of the Rabi frequency on the phonon number is the cornerstone for any conceivable filter mechanism.

Example matrix elements are shown in Fig.~\ref{fig:overview-sketch}~b). Note that the differences between the matrix elements become smaller for increasing $n$, thus larger discrimination times are required as the relevant part of the Hilbert space (i.e. if we want to have our map acting in the desired way for a higher range of initial states) increases.

\subsection{Fidelity Function}
Since the ion will initially be in a state with only a few motional excitations, we evaluate the fidelity of $U_m(T)$ (the imperfect implementation of $U_m$) only by its action on an $N$-dimensional subspace spanned by the basis states  $\{\vert \psi_{n}\rangle\}_{n=0\dots N-1}$, with $\vert\psi_{n}(0)\rangle=\vert \downarrow, n\rangle$. The map $U_m(T)$ evolves $\vert\psi_{n}(0)\rangle$ into $\vert\psi_{n}(T)\rangle=U_m(T) \vert\psi_{n}(0)\rangle$.
We choose the fidelity $F$ of $U_m(T)$ 
to be the product of the probability that the desired spin flip occurs for level $m$, and the averaged probability that no undesired spinflip occurs for $n\neq m$: 
\begin{eqnarray}\label{eq:fidelity}
F(m,N)=F_{\uparrow} F_{\downarrow}\,,
\end{eqnarray}
where
\begin{eqnarray}
F_{\uparrow}=\sum_{k=0}^{N-1}\vert \langle \uparrow, k\vert \psi_{m}(T)\rangle\vert^{2}\,,\\
F_{\downarrow}= \frac{1}{N-1}\sum_{n\neq m}\sum_{k=0}^{N-1}\vert \langle \downarrow, k\vert \psi_{n}(T)\rangle\vert^{2}\,.
\end{eqnarray}
So if $F=1$ it means that for any input state the spin up population after the mapping corresponds exactly to the population of $\ket{m}$ before the mapping. In other words the spin is fully flipped to $\ket{\uparrow}$ for initial state $\ket{m}$ and stays in $\ket{\downarrow}$ otherwise.

The maps $U_m$ will by design of $F(m,N)$ produce the desired spin operation only for those initial states that do not populate vibrational levels $|n\geq N\rangle$. Instead $U_m$ will act in an unspecified way on $|n\geq N\rangle$. 
Thus $N$ initial basis states are included when calculating the fidelity. Throughout the article we choose $N=10$, unless explicitely stated. We will then write $U_m^N$ to refer to the number of initial states included in the fidelity of the map.

\section{Simulation and Parameters}\label{sec:simulation}
The simulation of the unitary dynamics is carried out by truncating the Hilbert space, allowing for vibrational excitations of up to $|n=14\rangle$, i.\,e. the total dimension is 30. The Hamiltonian is mapped on this subspace and time dynamics can be simulated by diagonalizing the Hamiltonian at each time step. The validity of the truncation is verified by checking convergence of the simulation on a bigger subspace with dimension 40.

Realistic parameters for the ion light interaction are chosen with a maximal bare Rabi frequency $\Omega_0=2\pi \cdot 50\,$kHz, operation time $T=100-1000\,\mathrm{\mu s}$ with 1000 time steps that determine the grid for changes of the field amplitudes and thus allow us to represent pulse modulation frequencies up to about $(10-20)\cdot 2\pi/T$. The Lamb-Dicke parameter was set to $\eta=0.25$, the trap frequency $\omega_z=2\pi \cdot 1.4\,$MHz, so well in the regime $\omega_z\gg \Omega_0$ where resonant terms are dominant over offresonant terms. The parameters correspond to recent experiments \cite{Poschinger,Ziesel}.

\section{Optimization}\label{sec:optimization}
Now we are ready to engineer $U_m$ by simulating the system's time evolution and optimize the pulses via the Chopped Random Basis (CRAB) algorithm \cite{Doria,Caneva}.
The Rabi frequency ramps lead to the dimensionless pulses
\begin{eqnarray}
f^{(\alpha)}(t)=\Omega^{(\alpha)}(t)/\Omega_{0},
\end{eqnarray}
which are optimized so that the pulses vanish at zero and final time and $|f^{(\alpha)}(t)|\leq 1$, i.e. the maximum pulse height does not exceed the experimental constraint $\Omega_{0}$.

\subsection{Control with three radiation fields}
If we control the laser power on the carrier transition as well as on the red and blue sideband we have three (real) controls that are subject to CRAB.  %Figure \ref{fig:max-vs-T} shows the fidelities for the resulting operations $U_m$ as a function of operation time $T$.
If we allow also for the phases of the lasers to be controlled by CRAB we end up with 6 (real) controls, but it turns out that this does not lead to substantial improvement of the fidelities. Therefore in the following we restrict ourselves to real valued pulses. Note that we assume the lasers to be in phase at $t=0$, although an initial phase different will not cause a big effect \cite{RoosNJP}.

Optimization was done for selecting $m=0,1,2,3$. Fig.~\ref{fig:filtersketch} shows how the achieved mapping unitaries $U_m(T)$ transform initial spin down population into spin up population depending on the initial state $|\psi_{n}(0)\rangle$ along with the fidelities of the operations.

In an actual experimental realization, the main error source would result from an imperfect calibration of the Rabi frequencies. This error can be modelled as
\begin{eqnarray}
\tilde{\Omega}_0 f^{(\alpha)}(t)=\Omega_0(1+ \xi)f^{(\alpha)}(t)
\end{eqnarray}
where $\tilde{\Omega}_0 f^{(\alpha)}(t)$ are the actual Rabi frequencies while $\Omega_0f^{(\alpha)}(t)$ is the calculated optimal pulse.
The robustness of the results is thus characterized by the fidelity as a function of $\xi$. For $-0.01\leq \xi\leq0.01$ the fidelity does not differ by more than $1.3\,\%$ from its maximum value.

\begin{figure}
 \begin{minipage}{\textwidth}\flushleft
 \begin{overpic}[width=0.5\textwidth]{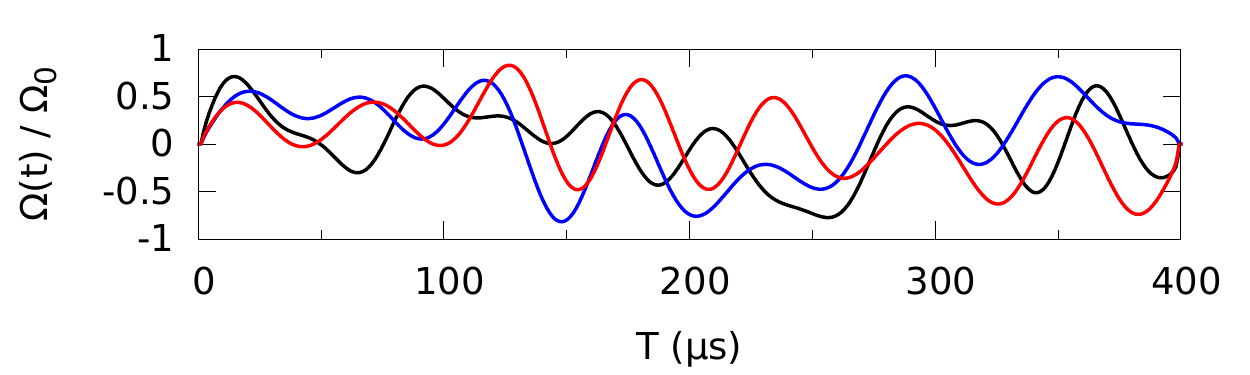}
  \put(0,27){a)}
 \end{overpic}
 \end{minipage}
\begin{minipage}{\textwidth}\flushleft
 \begin{overpic}[width=0.5\textwidth]{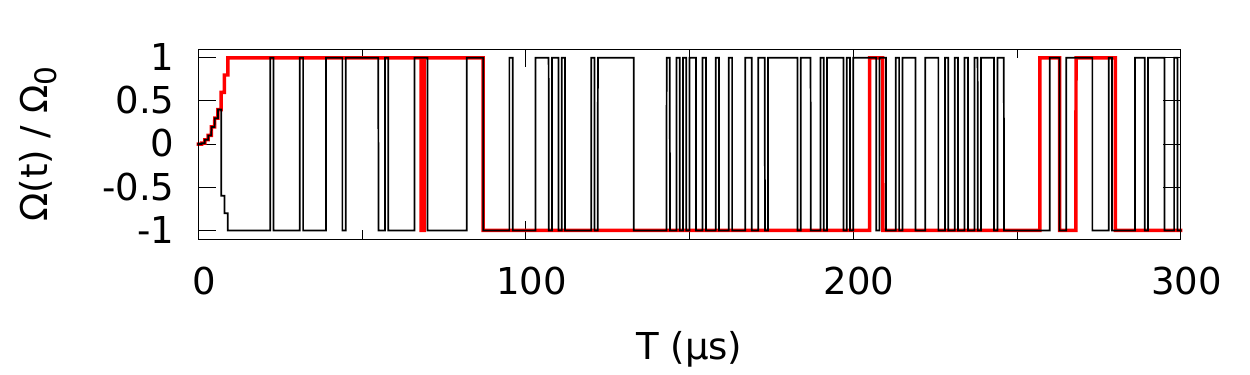}
  \put(0,27){b)}
 \end{overpic}
 \end{minipage}
\caption{Pulse shapes after optimization. Pulse height in units of $\Omega_0=2\pi\cdot 50\,$kHz. a) The optimal pulse for $m=1$, $T=400\,\mathrm{\mu}s$, where three radiation fields (black=carrier, red=red sideband, blue=blue sideband) are applied and all of them are modulated continuously in time. The transformation works on the harmonic oscillator levels $n=0,\dots 9$. b) Two fields (black=carrier, red=red sideband) are applied and 
the control is limited to phase switching by $\pi$ while the power is kept constant ($\Omega^{(\alpha)}(t)=\Omega_0$ or $\Omega^{(\alpha)}(t)=-\Omega_0$). The transformation operates on the harmonic oscillator levels $n=0,\dots 3$ for $m=0$, $T=300\,\mathrm{\mu}s$.}\label{fig:pulse}
\end{figure}

\subsection{Control with two radiation fields}
If we use only two driving fields the control task is more difficult to achieve. However, this is a very interesting problem not only because it requires less complexity when implemented in an experiment, but also since it corresponds to the systems analyzed e.g. by \cite{LawEberly,Rangan,Yuan,Mischuck}.
\begin{figure}[h]
 \begin{overpic}[width=0.5\textwidth]{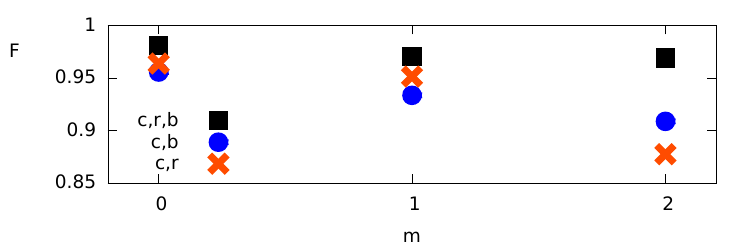}
 \end{overpic}
\caption{Optimization results for different control scenarios: fidelity $F(m,10)$ for three driving fields (black squares), driving only the carrier and the blue sideband transition (blue circles) and using only the carrier and the red sideband transition (red crosses) as a function of $m$.}\label{fig:compare}
\end{figure}
Figure \ref{fig:compare} compares the resulting fidelity $F(m,10)$ for three driving fields (black squares), with scenarios using the carrier and the blue sideband transition (blue circles) and using the carrier and the red sideband transition (red crosses). Both options work equally well.
In the following we consider one field resonant to the carrier and one field resonant to the red sideband transition since it is closer to the systems in Ref. \cite{LawEberly,Rangan,Yuan,Mischuck}.

\begin{figure}[h]
 \includegraphics[width=0.5\textwidth]{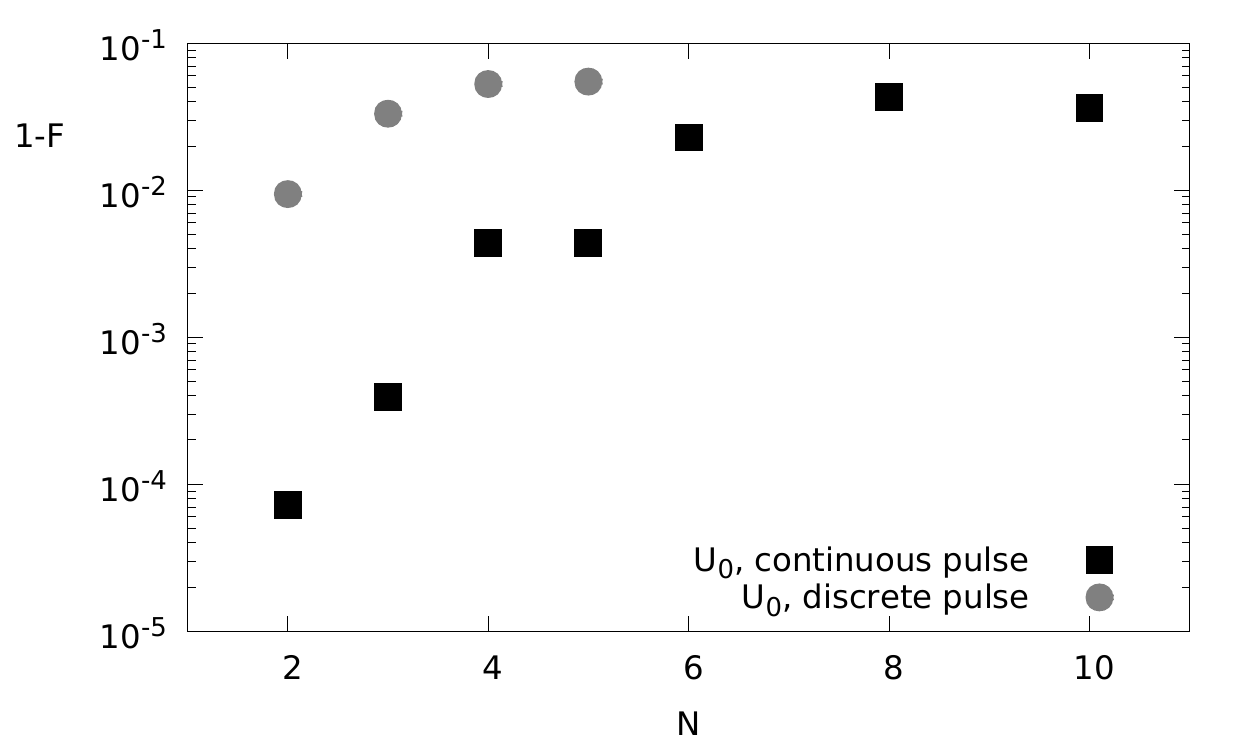}
\caption{Operation error $1-F$ as a function of $N$, the number of basis states considered. The results are for realizing $U_0^N$ in an operation time of $T=300\,\mathrm{\mu s}$ using carrier and red sideband. The black squares show the results allowing continuous modulation of the pulse power over time. The gray circles show the results obtained when only phase flips at constant power are allowed (compare Fig.~\ref{fig:pulse} for the pulse shape).}\label{fig:max-vs-Neval}
\end{figure}

So far we had $U_m$ operating in the desired way on the subspace of $N=10$, thus for up to 9 vibrational quanta. This was achieved by maximizing the control functional $F(m,N=10)$ that considers the time evolution of $N=10$ basis functions.
However, in a scenario with only few excitations it can be enough 
to distinguish e.g. $|n=0\rangle$ from $|n=1\rangle$ and $|n=2\rangle$ while it is known that there is no contribution from $|n\geq 3\rangle$. This requires a map that acts in a desired way only on $|\psi_n\rangle_{n=0,1,2}$ and can be achieved by maximizing $F(m,N=3)$, thus considering the time evolution of only $N=3$ basis functions. As presented in \cite{Mischuck}, the control task dramatically simplifies if fewer levels are included in the control functional. Fig.~\ref{fig:max-vs-Neval} shows the scaling of the fidelity $F(m,N)$ with $N$ for $m=0$ and $T=300\,\mathrm{\mu s}$ (black squares). As one can see, especially for small $N$ the fidelity is much higher than for intermediate and larger $N$.
One obvious reason, as mentioned already, is that we have to fix fewer parameters for smaller $N$ (e.g. the time evolution of 3 basis functions for $U_m^3$ as compared to 10 basis functions in the case of $U_m^{10}$) while the complexity of the optimization problems scales with the number of parameters \cite{LloydMontangero}.
Furthermore, for larger $N$ the transitions with higher $n$ are included, where the matrix elements $M_{n,n+\delta n}$ of the carrier and the sidebands become comparable to one
another and the scaling with $n$ becomes smaller and thus distinguishing between two vibrational levels becomes harder. The influence of this scaling of the transition strength was also studied in appendix \ref{sec:poincare} by considering the system as a set of Poincaré pointers.

A control strategy that is even simpler to realize experimentally but also more limited would be to keep the pulses on constant power and modulate the pulses just by phase flips ($\Omega^{(\alpha)}(t)=\Omega_0$ or $\Omega^{(\alpha)}(t)=-\Omega_0$). We call this a discrete pulse. The corresponding pulse is shown in Fig.~\ref{fig:pulse} b), while the resulting fidelities $F(m=0,N)$ are shown by gray circles in Fig.~\ref{fig:max-vs-Neval}. We present results for $N=2,\dots 5$ since the restrictions on the pulse make the control task much more challenging and if we include more levels the fidelity drops. Still we get good fidelities for small $N$.

\section{Conclusions}
We have obtained control fields for phonon-selective flipping of the internal state of a trapped ion by application of the CRAB algorithm. The optimization yields solutions with fidelities of better than 95\,\% where total times $\leq 500\mathrm{\mu s}$ are required for the mapping transform laser pulse.
This bears interesting applications for the quantum state reconstruction of the motional mode, and for the measuring of the work distribution pertaining to quantum processes. This in turn opens up perspectives in the emerging field of quantum thermodynamics \cite{An,QThD}. 
An important feature is that we do not specify the final state of the harmonic oscillator degree of freedom. This is an algebraic reduction of the degrees of freedom to be fixed by optimal control and has been shown to lead to significantly better control performance compared to full gate optimization in a slightly different context \cite{Mueller}.
Another interesting aspect of the mapping transform is that for only one laser it reduces to the Poincaré recurrence time problem. This analogy is studied in appendix \ref{sec:poincare}.
While we focus on trapped ions the results are also relevant for systems of atoms in cavities and superconducting qubits.

\begin{acknowledgments}
The authors acknowledge support from SFB/TRR21, Q.com, and the EU project SIQS and we thank the bwUniCluster~\cite{Cluster} for the computational resources.
\end{acknowledgments}
\appendix

\section{Hamiltonian}
First we consider the contribution of a single radiation field to the Hamiltonian. We label the fields by $\alpha=c,b,r$ and characterize them by the bare Rabi frequency $\Omega_0^{(\alpha)}$ and the detuning from the carrier transition $\delta^{(\alpha)}_c=\omega_\alpha-\omega_c$ ($\omega_c$ the carrier transition and $\omega_\alpha$ the frequency of the field). To have the fields on resonance with the carrier ($\Omega_0^{(c)}$) and the blue and red sidebands ($\Omega_0^{(b)}$ and $\Omega_0^{(r)}$, respectively) we have to choose $\delta^{(c)}_c=0$, $\delta^{(b)}_c=\omega_z$, and 
$\delta^{(r)}_c=-\omega_z$, where the trap frequency $\omega_z$ is the energy of one phononic excitation.

 The contribution of one of these fields to the total Hamiltonian in the interaction picture reads
\begin{eqnarray}\label{eq:ham-1field}
 H^{(\alpha)}(t)&=&\vert \uparrow \rangle\langle \downarrow \vert\otimes\sum_{n=0}\bigg( \frac{\Omega^{(\alpha)}_{n,n}(t)}{2}\mathrm{e}^{i\delta_c^{(\alpha)} t}\vert n \rangle\langle n\vert + h.c.\nonumber\\
&+&\frac{\Omega^{(\alpha)}_{n,n+1}(t)}{2}\mathrm{e}^{i(\delta_c^{(\alpha)} -\omega_z)t}\vert n+1 \rangle\langle n\vert + h.c.\nonumber\\
&+&\frac{\Omega^{(\alpha)}_{n,n-1}(t)}{2}\mathrm{e}^{i(\delta_c^{(\alpha)} +\omega_z)t}\vert n-1 \rangle\langle n\vert + h.c.\bigg)\,,\nonumber\\
\end{eqnarray}
thereby neglecting contributions that connect phonon numbers that differ by more than one.  The Rabi frequencies depend on the phonon number $n$ and the Lamb-Dicke factor $\eta$ as 
\begin{eqnarray}
 \Omega^{(\alpha)}_{n,n+\delta n}(t)=\Omega^{(\alpha)}(t) M_{n,n+\delta n}(\eta)
\end{eqnarray}
where the matrix elements are
\begin{eqnarray}
 M_{n,n}(\eta)=\mathrm{e}^{-\eta^2/2}L_n^0(\eta^2)\\
 M_{n,n+1}(\eta)=i \mathrm{e}^{-\eta^2/2}\eta \sqrt\frac{1}{n+1}L_{n}^1(\eta^2)\\
 M_{n,n-1}(\eta)=i \mathrm{e}^{-\eta^2/2}\eta\sqrt\frac{1}{ n}L_{n-1}^1(\eta^2)\,.
\end{eqnarray}
The functions $L_n^k$ are the standard Laguerre polynomials.
We assume that the amplitudes of the drive fields $\Omega^{(\alpha)}(t)$ can be varied arbitrarily over time.

The total Hamiltonian then reads
\begin{eqnarray}
 H(t)=H^{(c)}(t) + H^{(b)}(t) +H^{(r)}(t)\,.
\end{eqnarray}
The fact that the Rabi frequencies depend on the vibrational quantum number $n$ enables our desired spin flipping mechanism.

\section{Offresonant terms -- Stark shift}
As we can see in equation (\ref{eq:ham-1field}) each laser contributes not only the resonant term but also non-resonant terms. We neglect terms rotating at $2\omega_z$, and terms which scale with a $|\delta n|> 1$ matrix element. The remaining four terms are:
\begin{eqnarray}
\sum_{n=0}\frac{\Omega^{(c)}_{n,n\pm 1}(t)}{2}\mathrm{e}^{\mp i\omega_z t}\vert \uparrow \rangle\langle \downarrow \vert\otimes\vert n\pm 1 \rangle\langle n\vert + h.c.\,,\\
\sum_{n=0}\frac{\Omega^{(b)}_{n,n}(t)}{2}\mathrm{e}^{i\omega_z t}\vert \uparrow \rangle\langle \downarrow \vert\otimes\vert n \rangle\langle n\vert + h.c.\qquad\mathrm{and}\\
\sum_{n=0}\frac{\Omega^{(r)}_{n,n}(t)}{2}\mathrm{e}^{-i\omega_z t}\vert \uparrow \rangle\langle \downarrow \vert\otimes\vert n \rangle\langle n\vert + h.c. \,,
\end{eqnarray}
Together with the resonant terms this corresponds to equation (\ref{eq:Ham-allfields}). All other offresonant terms are neglected and even these four terms are small since $\omega_z\gg \Omega^{(\alpha)}$.

\section{The Poincaré Recurrence Time}\label{sec:poincare}
The desired spin mapping operation $U_m$ could be accomplished also without sophisticated control pulses but with a constant pulse on the e.g. blue motional sideband. This is related to Poincaré's \cite{Poincare} recurrence time theorem. The theorem states that a conservative, finite, closed system returns arbitrarily close to its initial state after the \textit{Poincaré recurrence time}. Hemmer et al. \cite{Hemmer} showed that for a chain of $N$ harmonic oscillators this boils down to the problem of $N$ pointers, all rotating with a certain frequency $\omega_{n}$, where these frequencies are incommensurate. Thus, they can calculate the Poincaré recurrence time from the probability of having all pointers directing in the same direction up to a directional error $\Delta \phi_n$.

\begin{figure}
\includegraphics[width=0.5\textwidth]{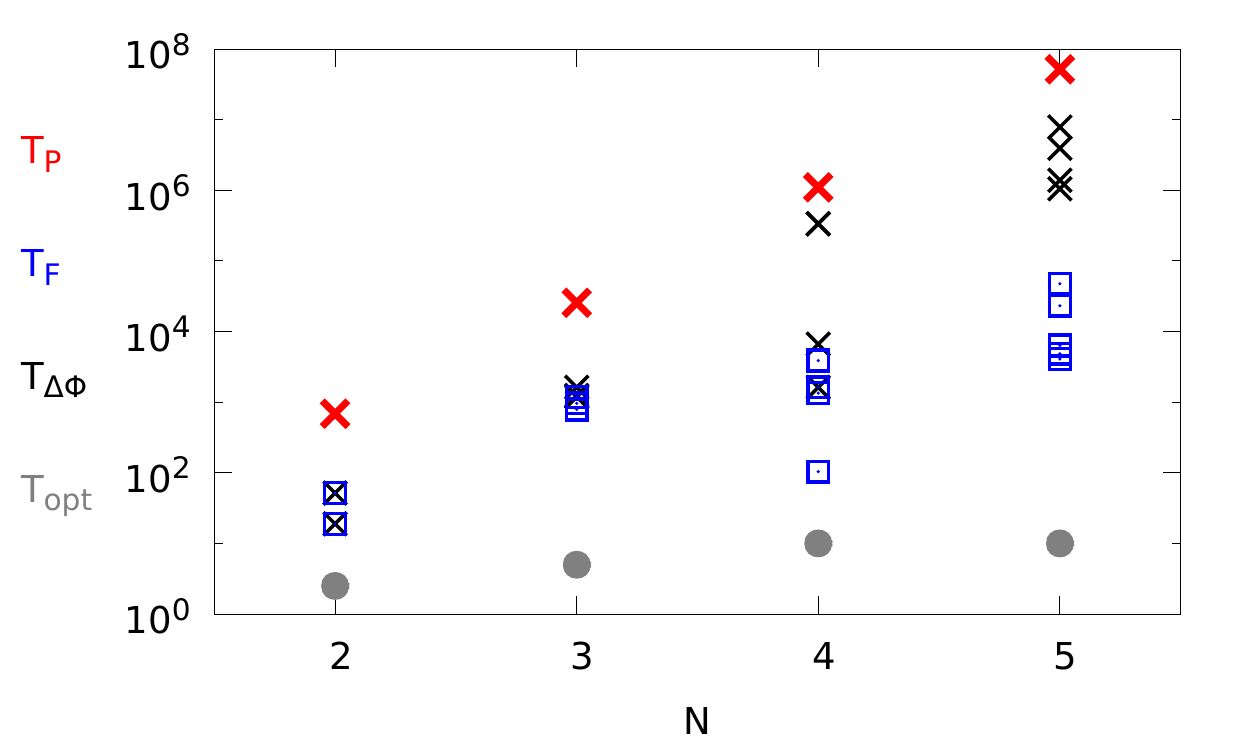}
\caption{Waiting times in units of $1/\Omega^{(b)}_0$. Theoretical Poincaré recurrence times $T_P$ (red crosses) and required waiting times found by numerical simulations (black crosses and blue squares) are plotted for different numbers $N$ of included phonon excitations. The results were obtained for $\Omega^{(b)}_{0}=2\pi\cdot 50\,$kHz and $\varepsilon=0.02$. Blue squares are waiting times $T_{F}$ for crossing the error threshold (equation \ref{eq:poincare-epsilon}), black crosses are waiting times $T_{\Delta\phi}$ for constant $\Delta\phi$ (equation \ref{eq:poincare-const-phi}). The gray circles show $T_{\mathrm{opt}}$, the operation time needed for the optimized maps $U_0(N)$ with carrier and red sideband presented in Fig.~\ref{fig:max-vs-Neval}.}\label{fig:Poincare}
\end{figure}
If we try to solve our spin mapping problem for the simple case of driving the blue sideband (where we can choose the amplitude to be constant without loss of generality) on the subset of $n=0,..,N-1$, this corresponds to a set of $N$ rotating pointers, that in the end all have to point in a certain \emph{given} direction (given by the desired spin flip described by $U_m$) up to an error $\Delta \phi_{n}$. The frequencies of the pointers are $\omega_n=\Omega^{(b)}_{n,n+1}$. In the Lamb-Dicke regime $\eta^2 n\ll 1$, $\Omega^{(b)}_{n,n+1}\propto \sqrt{n+1}$, from which we see that the frequencies are actually incommensurate.
Following \cite{Hemmer}, the theoretical waiting time $T_p$ after which $U_m$ is generated up to some specified error is
\begin{eqnarray}
\frac{1}{T_P}=\frac{1}{(2\pi)^N}\sum_{n=0}^{N-1}\Omega^{(b)}_{n,n+1}\prod_{\tiny\begin{matrix}k=0\\ k\neq n\end{matrix}}^N\Delta\phi_{k}\,.
\end{eqnarray}
The errors due to $\Delta\phi_k\neq 0$ correspond to a fidelity of
\begin{eqnarray}
F_{\uparrow}=\cos^2\Delta\phi_{m}\\
F_{\downarrow}= \frac{1}{N-1}\sum_{n\neq m}\cos^2\Delta\phi_{n}\\
F=F_{\uparrow}F_{\downarrow}\approx 1- \Delta\phi_{m}^2-\sum_{n\neq m}\frac{\Delta\phi_{n}^2}{N-1}\,,
\end{eqnarray}
where the approximation holds for $\Delta\phi_{n}\ll 1$. For a given maximal error $\varepsilon=1-F$ thus we have to find $\Delta\phi_{n}$ such that
\begin{eqnarray}
\Delta\phi_{m}^2+\sum_{n\neq m}\frac{\Delta\phi_{n}^2}{N-1}\leq \varepsilon\label{eq:poincare-epsilon}\,,
\end{eqnarray}
while under the further constraint of $\Delta\phi_n$ not depending on $n$ the condition on $1-F$ translates into
\begin{eqnarray}
 \Delta\phi_n\leq \sqrt{\varepsilon/2}\label{eq:poincare-const-phi}\quad n=0,\dots N-1\,.
\end{eqnarray}
We calculate $T_P$ with this latter condition (i.e. setting $\Delta\phi_n=\sqrt{\varepsilon/2}$)
\begin{eqnarray}
 \frac 1{T_P}=\frac{(\frac \varepsilon 2)^{\frac{N-1}{2}}}{(2\pi)^N} \sum_{n=0}^{N-1}\Omega_{n,n+1}\,,
\end{eqnarray}
where we set the operation error to $\varepsilon=0.02$.

This theoretical Poincaré recurrence time can be compared to waiting times obtained by numerical time evolution: we evolve the system under constant $\Omega^{(b)}(t)=2\pi\cdot 50\,$kHz and stop when our conditions on the phase deviations or fidelity holds. We call $T_{F}$ the time point, where in the numerical time evolution equation (\ref{eq:poincare-epsilon}) holds for the first time and likewise $T_{\Delta\phi}$ the time point where equation (\ref{eq:poincare-const-phi}) holds for the first time. Figure \ref{fig:Poincare} shows $T_P$ (red crosses) together with $T_{F}$ (blue squares) and $T_{\Delta\phi}$ (black crosses) for $N=2,\dots 5$. For the numerical values different points for same $N$ correspond to $m=0,\dots N-1$.
The Poincaré waiting time is an approximation for the stricter condition
of equation (\ref{eq:poincare-const-phi}) and thus closer to $T_{\Delta\phi}$. As seen in the figure indeed the condition on $\Delta\phi_n$ is stricter, meaning longer waiting times ($T_{\Delta\phi}\geq T_{F}$).

These waiting times are now compared to the control scenario of two radiation fields (carrier plus red sideband) and the pulse shapes are optimized for each $N$ as in Fig.~\ref{fig:max-vs-Neval}.
The gray circles show $T_{\mathrm{opt}}$, the operation time required to achieve $U_0(N)$ up to an error $\varepsilon=0.02$. While the presented values are too few to make general statements about the scaling of $T$ with $N$ it is clear that by optimization one can gain a speed-up of several orders of magnitude.


\begin{thebibliography}{10}
\bibitem{Cluster}This work was performed on the computational resource bwUniCluster funded by the Ministry of Science, Research and Arts and the Universities of the State of Baden-Württemberg, Germany, within the framework program bwHPC.
\end{thebibliography}
\end{document}